# Implementation of Two Way Free Space Quantum Key Distribution

**M. F. Abdul Khir**[1,3], **M. N. Mohd Zain**[3], **Suryadi**[2], **S. Saharudin**[3], **S. Shaari**[1]

[1] Photonic Lab, IMEN, Universiti Kebangsaan Malaysia, 43400 UKM Bangi, Malaysia
[2] Faculty of Science, International Islamic University of Malaysia (IIUM), Jalan Bukit Istana, 25200 Kuantan, Pahang, Malaysia
[3] Photonic Lab, Micro Systems and MEMS, MIMOS Berhad, Technology Park Malaysia, 57000 Kuala Lumpur, Malaysia

**Abstract:**

We report an implementation over free space medium of a two way four states quantum key distribution (QKD) protocol namely the LM05. The fully automated setup demonstrated a secure key generation rate of 3.54 kbits per second and quantum bit error rate (QBER) of 3.34% at mean photon number ($\mu$) = 0.15. The maximum tolerable channel loss for secure key generation considering Photon Number Splitting (PNS) attack, was 5.68 [dB]. The result successfully demonstrated the feasibility of a two way QKD protocol implementation over free space medium.

## Introduction

The last two decades has seen a remarkable development in the field of Quantum Cryptography (QC), broadened from quantum physics and information theory to research activities in engineering aspect for practical implementation. The attractions of this technique lies in the ability of distributing keys between two parties unconditionally without having to rely on third parties, which is something not achievable with the classical way. Rather than relying on mathematical difficulties as in present conventional asymmetric based cryptography schemes, the security of the shared key in QC relies on the law of physics, suggesting a very reliable foundation. However, as QC only solves the key distribution problem, it can only be used to complement the standard symmetrical cryptosystems [1]. This led to a more precise name, Quantum Key Distribution (QKD).

Since the introduction of BB84 protocol in 1984 [2], QKD implementations have been demonstrated over optical fiber [3-6] and free space [7-9] medium. Recent experiments surpassing 100 km distant for free space [10] and 200 km for optical fiber [11] had been demonstrated thanks to the decoy state method. Efforts to realize a QKD into current network infrastructure can be seen for example in [12,13].

In recent years, several research groups have proposed new QKD protocols of which include variants of the two way protocols [14,15,16]. Although several experiments have been conducted to realize the proposed protocols such as in [17-20], as compared to the rest in the group, the LM05 protocol has undergone quite a significant development in its realization. Several experimental works on near infrared wavelength were carried out by Lucamarini et al to realize the LM05 protocol [17,18]. In recent experiment, a complete system at telecommunication wavelength was successfully realized by Kumar et al [21]. Their system used the plug and play type and had shown an almost equivalent secure key generation rate to the one developed by [22] for BB84 protocol. However, a complete system that runs over a free space medium has yet to be realized.

In this work, we demonstrate experimentally an implementation of LM05 protocol over free space medium. To the best of our knowledge this is the first time such fully automated LM05 protocol implementation over free space medium is realized. This setup allows to a certain extend an experimental investigation of the capability of a practical free space based LM05 protocol implementation in producing secure keys. Following [21], we simulate the effects of control mode using a beam splitter. As such, this letter is organized as follows. Section two introduces the LM05 protocol. Section three explains the experimental setup, followed by results and discussion in section four. Section five conclude and suggest future works.

**2. Protocol**

Let us briefly discuss the encoding mode of the LM05 QKD protocol realized with polarization states of single photon (refer to Fig. 1). Similar to BB84, two sets of non-orthogonal basis states of single photon are used. However, contrary to BB84, it is Bob who would initiate the quantum key distribution process by preparing random sequence of bit using one of the four linear polarization states and sends them to Alice who would then use the polarized photon to encode a bit of information. Alice would choose to either encode a logical 0 by applying identity operator (I) or encode a logical 1 by applying unitary transformation iY. Alice then sends the qubit back to Bob who will measure the qubit in the same basis he used to prepare it. Bob then decodes the returning qubit into a sequence of bit that forms the raw key. At this point, it is not necessary for Alice and Bob to proceed with basis reconciliation as found in BB84 protocol due to the deterministic nature of the protocol. For this reason, ideally the raw key is the sifted key. They then proceed for error correction and privacy amplification which results in a set of identical secure keys at both sides.

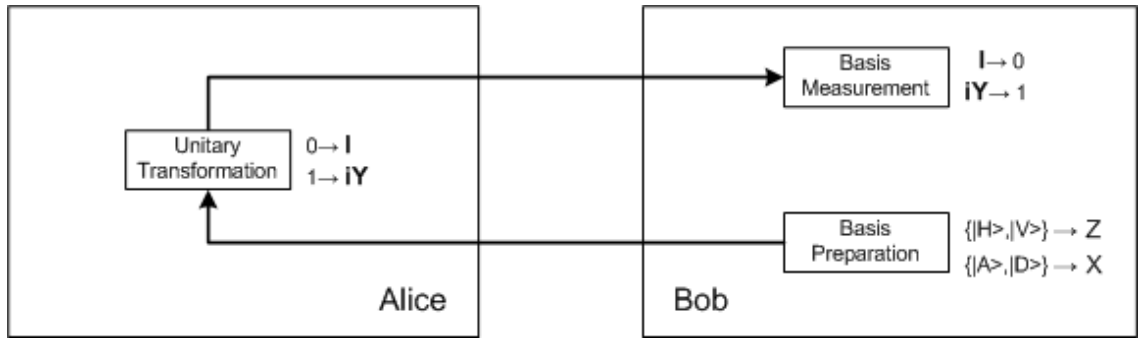

Fig. 1. The encoding mode of LM05 protocol. Bob prepares qubit with random bases (Z or X) and sends to Alice who would either encodes with identity operator (I) or unitary transformation (iY) and returns to Bob. Bob measures the qubit using same basis he used to prepare.

## 3. Experimental Setup

The schematic of our experimental setup is depicted in Fig. 2. A pair of computer programs based on LabVIEW 8.5 was developed at Alice and Bob to control and synchronize the whole active optical components via means of field-programmable gate array (FPGA). The FPGA is a 40 MHz Reconfigurable I/O module of National Instruments (PXI-7833R) at both Bob and Alice). We configured the FPGA pair down to 0.725MHz repetition rate to match the limitation of our Pockels cell (~ 1 MHz). All active optical components are connected to the FPGA using coaxial cables.

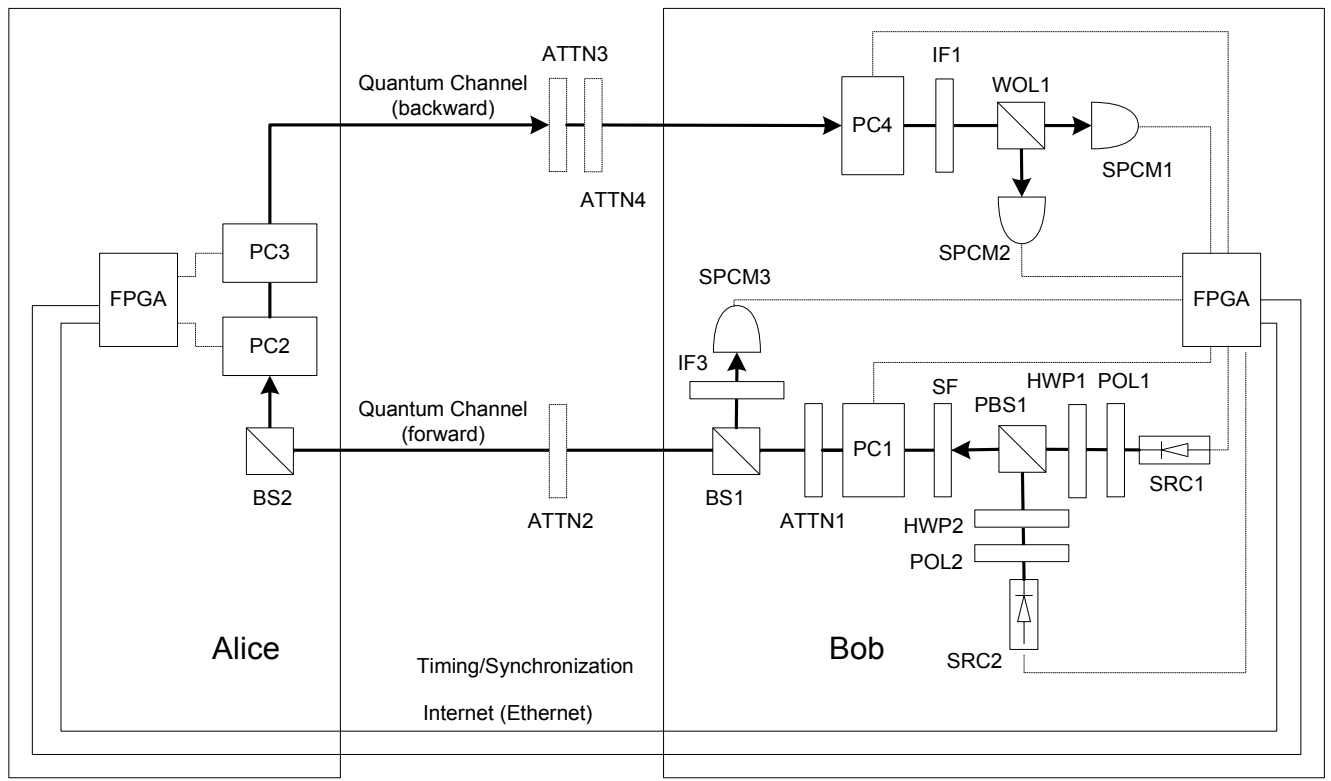

Fig. 2. The LM05 free space QKD experimental setup consists of SRC1, photon source; SRC2, photon source ; PBS1, polarization beam splitter; SF, spatial filter; PC1, first Pockels cell; ATTN1,ATTN4, variable attenuator; BS1,BS2 50/50 beam splitter; ATTN2, ATTN3 attenuator; PC2, second Pockels cell; PC3, third Pockels cell; PC4, Fourth Pockels cell; IF1, IF2, interference filter; WOL1, Wollaston Prism; SPCM1, H & D detector; SPCM2, V & A detector

The photon source consists of two pulse lasers (SRC1 and SRC2) from Coherent Connection CUBE with wavelength at 785 nm. Either SRC1 or SRC2 is randomly triggered one at a time to emit horizontally and vertically polarized pulse respectively. The optical pulse from both photon sources is separately attenuated in order to ensure that they carry same intensity before being combined into the same optical path at polarization beam splitter (PBS1). To improve the spatial mode quality, the optical pulse is passed through a spatial filter which is a four meters single mode three stage polarization controller from Fiber Control with operating wavelength 780~900nm. Optical pulse from either source then proceed through a high speed Rubidium Titanyle Phosphate, RbTiOPO4 (RTP) non-linear material; Leysop, RTP3-20 (PC1) to produce the four linear polarization states. The PC1 orientation and voltage are set so that the horizontally or vertically polarized pulse from SRC1 or SRC2 is transformed to either Diagonal or Anti-diagonal respectively at every triggering event. That is to say, not triggering the PC1 prepares state of rectilinear basis (|H>, |V>) while triggering the PC1 prepares state of diagonal basis (|A>, |D>). The combination of randomly triggered SRC1, SRC2 and PC1 produces the four linear polarization states. The optical pulse is further attenuated at variable attenuator (ATTN1) for intended mean photon number ($\mu$), tracked with a 50/50 beam splitter (BS1) and a single photon counting module (SPCM3) before being launched to Alice via the quantum channel. The quantum channel is a free space medium of several tens of centimeters length and the public channel is an Ethernet connection for public discussion.

The detector part consists of an active basis selector which is a high speed Rubidium Titanyle Phosphate, RbTiOPO4 (RTP) non-linear material (PC4; Leysop, RTP3-20), a Wollaston prism (WOL1) and a pair of single photon counting modules (SPCM1 and SPCM2). The function of PC4 is similar to PC1 in terms of switching between the two measurement basis. It is only triggered when PC1 is triggered where it counter rotates the Diagonal and Anti-diagonal basis for measurement by the WOL1 and the single photon counting module pair. That is to say, when PC4 is not triggered, the whole detector part is set to measure the |H>,|>V basis while if triggered, the whole detector part is set to measure the |D>,|A> basis. As such, the operation of active basis selector (PC4) deterministically sets the measurement basis to the same basis the

photon was prepared before launched to Alice. All incoming photons are filtered using an interference filter (IF) to reduce noise before entering WOL1 where they were collected into two multimode fibers and proceed to either SPCM1 or SPCM2 for detection. The single photon counting module pair (SPCM1 and SPCM2) is a Perkin Elmer SPCM-AQR-16-PC with 55% efficiency. We set the detection window for 25 ns, the smallest possible with our 40 MHz FPGA clock to reduce contribution from background noise.

Alice optical setup simply consists of an automated polarization flipper and a 50/50 beam splitter (BS2). The function of BS2 is to simulate the effect of control mode which is absent in this setup. Such similar arrangement was the approach adopted by Kumar et al in [21]. The flipper is constructed using two serially aligned Pockels cells (PC2 and PC3) which orthogonally rotate any four linear input polarization states. Since the polarization state of incoming photon is not priory known to Alice, it is important that Alice must be able to orthogonally rotate any of the four linear polarization states by just triggering the flipper. Its operation was discussed in [25]. Based on random bit obtained from pseudo random number generator in FPGA program, Alice would either trigger the flipper to encode bit 1 or not trigger to encode bit 0 and returns the encoded photon to Bob.

It is obvious until now that this experimental setup involves many Pockels cells which results in more active polarization transformations than what is found in many QKD implementations reported so far and hence deserves a separate section to detail out the explanation. As such, we refer the readers to Appendix A for description on active polarization transformations involved in this setup.

## 4. Results and Discussion

It is known to date that the security of two way QKD protocols including LM05 is studied in the context of selected types of attack [27]. Here, we would like to consider the case for Photon Number Splitting (PNS) attack which is very much relevant to a practical QKD implementation. The security of LM05 protocol against PNS attack was analyzed in detail by Lucamarini et al in [18] and further in [28]. As such, let us briefly review the model used in this setup following the one in [18].

### A. Model

We assume photon source is a laser source emitting weak coherent state where the number of photon in each pulse follows Poisson distribution with mean photon number (μ). The probability of finding i photons in a pulse $P(i, \mu)$ is given by :

$$P(i,\mu) = \frac{\mu^i}{i!} e^{-\mu} \quad (1)$$

Assuming the intrinsic loss of the system consist of loss in Bob ($lB$) and loss in Alice ($lA$) (measured in dB), the transmission in Bob $t_B$ and Alice $t_A$ is given by $t_B = 10^{-\left(\frac{lB}{10}\right)}$ and $t_A = 10^{-\left(\frac{lA}{10}\right)}$ respectively. The internal transmission of the system including the detection probability is given by

$$\eta_{Bob} = t_A t_B \eta_{Det} \quad (2)$$

where $\eta_{Det}$ is quantum efficiency of Bob's detector.

The channel transmission $t_C$ is given by

$$t_C = 10^{-\left(\frac{2lC}{10}\right)} \quad (3)$$

Where $lC$ is one way loss of the quantum channel between Bob and Alice. The factor of two is for two way channel loss in this protocol (we assume same loss for forward and backward channel).

The overall transmission ($\eta$) is then given as

$$\eta = t_C \eta_{Bob} \tag{4}$$

From here signal detection probability ($P_{Signal}$) is given by

$$P_{Signal} = 1 - e^{-\eta\mu} \tag{5}$$

The overall dark count probability ($P_{Dark}$) is given by

$$P_{Dark} = 2db \tag{6}$$

where $db$ is the dark count per detection window. The factor of two is due to the two detectors in Bob which are the sources of dark count in our setup.

The overall detection probability which includes signal and dark counts is given as.:

$$P_{All} = P_{Signal} + P_{Dark} - P_{Signal}P_{Dark} \cong P_{Signal} + P_{Dark} \tag{7}$$

where $P_{Signal}P_{Dark}$ represent coincidence of detection between signal and dark count and can be neglected in experiment.

The overall Quantum Bit Error Rate (QBER) denoted as $E_{All}$ is adopted from [23] and is given by

$$E_{All} = \frac{e_0 P_{Dark} + e_{detector} P_{Signal}}{P_{All}} \tag{8}$$

where $e_0$ is erroneous detection due to background noise equal to $\frac{1}{2}$ and $e_{detector}$ is probability of erroneous detection from signals.

The secure key generation rate against PNS attack is given by [18] as :

$$R_{PNS} = P_{All}[\beta(1-\tau') - f_{casc}H(e) \qquad \tau' = \tau(e/\beta) \tag{9}$$

where

$\beta$ : Security parameter defined as $\beta = \frac{P_{All} - P'}{P_{All}}$ (10)

where

$$P' = 1 - (1 + \mu + \frac{\mu^2}{2} + \frac{1}{2}\frac{\mu^3}{6})e^{-\mu}.$$

$f_{casc}$ : Error correction efficiency based on cascade protocol

$\tau$ : Fraction of the key to be discarded during privacy amplification.

$\tau(e) = \log_2(1 + 4e - 4e^2)$ if $e < \frac{1}{2}$ and $\tau(e) = 1$ if $e \geq \frac{1}{2}$

where $e$ is the QBER.

$H(e)$ : Binary Shannon entropy.

$$H(e) = -e\log_2 e - (1-e)\log_2(1-e)$$

From here, we first obtained intrinsic parameters from our experimental setup and numerically simulate the optimal mu for every distance and find the maximum secure distance represented in channel loss ($lC$). The intrinsic loss was measured to be 6.426 [dB] in Alice and 3.781 [dB] in Bob. The high loss in Alice was mainly coming from the 50/50 beam splitter (3.378 [dB]) for simulating the effects of control mode and from the two Pockels cells (PC2 and PC3) that forms the flipper (1.524 [dB]). The loss in Bob came from the Pockels cell (PC4) and imperfect fiber coupling which are recorded as 1.06 [dB] and 3.219 [dB] respectively. As such, total intrinsic loss of our setup is 10.705 dB.

We further obtained $e_{detector} = 0.033$ and $P_{Dark} = 4.276 \times 10^{-6}$. The $e_{detector}$ is the noise related to the setup that is the optical alignment and the electronic components and is actually the QBER value when dark count is negligible. It is obtained by sending pulses with high mean photon number ($\mu$) and is almost consistent if the channel loss is not too high. $P_{Dark}$ is dark count probability in within the 25 ns detection window used in this setup when no pulse is sent. The double count was found to be very small and is neglected this experiment. We then conduct the experiment starting with characterization of mean photon number (μ) and followed by determination of maximum secure distance.

**B. Mean photon number (μ).**

An optimal mean photon number (μ) is given by the ratio of counts over triggers in a frame [19]. It is usually measured before launching into quantum channel. A same result could also be obtained if one use back end detectors (denoted as SPCM1and SPCM2 in Fig 2) by considering intrinsic loss and channel loss in between such as shown in [19]. In this setup, after a simple calibration to ensure same results within experimental tolerance, each μ is measured using the μ tracker (SPCM3 in Fig 2) such as found in [10,21].

In order to study the effects of μ over QBER, we fixed the attenuation at both forward and backward channel using ND filter (ATTN2 and ATTN3 in Fig. 2) and variable attenuator (ATTN4 in Fig 2.) which correspond to 1.14[dB] of one way channel loss and perform the experiment over several different values of μ < 1.7 .

For each experimental runs, we recorded the resulted QBER, obtained by evaluating the ratio of wrong bits over total detected bits. The results are depicted in Fig 3 which shows the QBER for a given µ. It is obvious that at the region outside µ < 0.1, QBER remains stable below 4%. At this region, the sources of QBER are mainly from the imperfect setting of the setup with dark count almost negligible. This value amounts close to the value of previously mentioned $e_{detector}$ at 0.033 and is mainly contributed by the three Pockels cells (two in Alice and one in Bob) where each polarization encoded photon has to travel through in its way to the detector. Imperfect polarization transformation had reduced the polarization contrast which as a result increased the probability of reaching wrong detector. When µ decreases surpassing 0.1, we can see that the QBER started to fluctuate. This is due to less photon flux reaching the detectors while dark count remains constant. We noted that it is important to keep the stable QBER value as low as possible since the controlling of Eves relies on this value while there is no means of differentiating the contribution of eavesdropper or imperfections of the setup. In the next sub-section, we study the effects of varying channel loss and further determine the maximum secure distance against PNS attack of this setup.

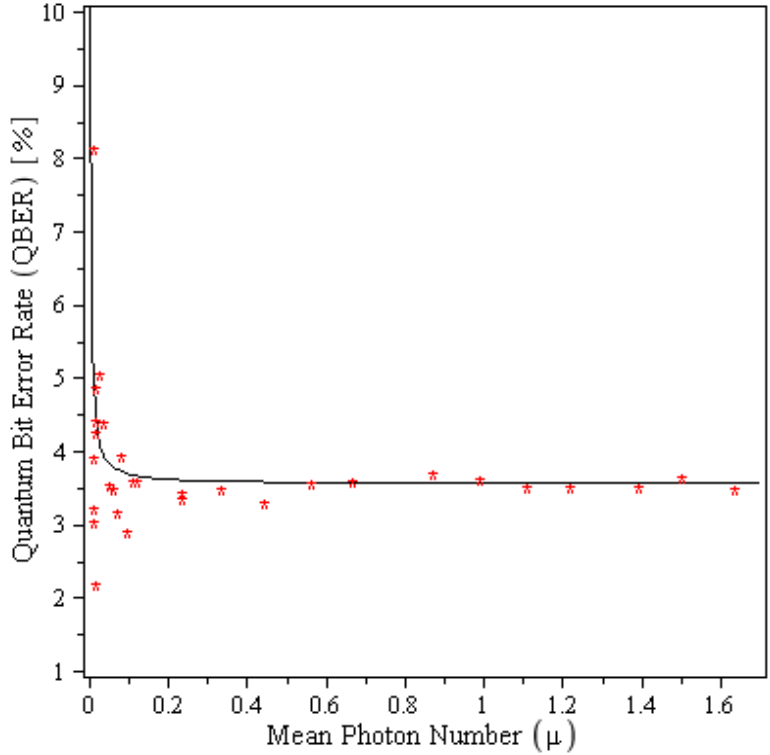

Fig. 3. QBER value with varying mean photon number ($\mu$) corresponding to channel loss = 1.14[dB]. The theoretical line is plotted according to Eq [8] with $e_{detector} = 0.033$ , $P_{Dark} = 8.552 \times 10^{-6}$.

**C. Secure distance.**

It is known that apart from system intrinsic loss, the real implementation of a free space QKD system suffers from environmental effects such as bad weather condition or even pollution. This is evident from recent field trials such as in [7~10]. All these effects eventually sum up to total channel loss. As such, in this experiment, we simulate the channel loss by varying attenuation level of the quantum channel as reported in several previous experiments [21, 24][26]. We fixed the value of μ to 0.15 which is not far from optimal as suggested by numerical simulation and slowly vary the quantum channel loss (denoted as $lC$ in section A) to investigate the maximum secure distance of our setup. Several low ND filters were inserted at the forward channel and backward channel denoted as ATTN2 and ATTN3 combined with ATTN4 respectively in Fig. 2. It is difficult to precisely equalize the attenuation value of ATTN3 and ATTN4 to ATTN2 and hence the one way channel loss reported here is actually the average of the two attenuation values. For each experimental run, we record the raw key and the corresponding QBER. Interestingly, it is found that the QBER value has not been affected much by the increase in the channel loss

We then calculate the corresponding secure key generation rate ($R_{PNS}$) for each varying distance using Eq 9 with error correction efficiency $f_{casc} = 1.22$. The result is depicted in Fig. 4 where we can easily spot that the maximum secure distance against PNS attack of this system is less than 6 [dB]. The maximum secure distance is obtained by observing the channel loss at the last point before secure key generation rate hits zero. In our case, with mean photon number ($\mu$) ≅ 0.15, as shown in Fig 4, the maximum secure distance is 5.68 [dB] producing 11.40 bit/sec secure key at pulse repetition rate of 0.725 MHz of this system. We verified that at 5.99 [dB] the security parameter is already negative which means that the system is not any more secure against PNS attack. The system was stable even for more than an hour of continuous operation.

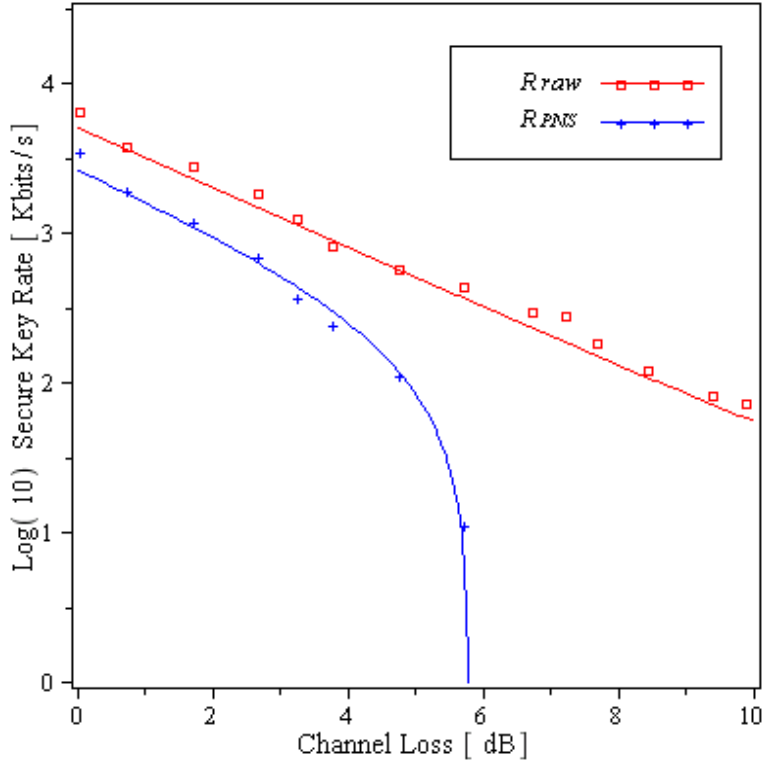

Fig. 4. Raw key rate $(R_{raw})$ and secure key generation rate $(R_{PNS})$ against channel loss [dB]. The secure key is calculated according to Eq (9).

## B. Conclusion and Future Works

We have successfully implemented the first fully automated QKD system based on a four states two way quantum key distribution protocol namely the LM05 over free space medium. Despite uses of many active components for polarization encoding, the system recorded a stable QBER as low as 3.44% with maximum tolerable channel loss for secure key generation at 5.68 [dB]. This strongly indicates the feasibility of two way QKD protocol over free space medium.

It is known that the secure key generation rate against PNS attack and the maximum secure distance can be improved by almost double if one employs decoy state method such as reported for LM05 protocol in [27, 29] . It is then interesting to experimentally realize the proposed scheme. In our setup, it would require at least a set of identical laser sources to produce the minimum additional four states for the decoy scheme. This highlights possible immediate extension of this setup.

## ACKNOWLEDGEMENTS

The authors would like to thank Youn-Chang Jeong and Yong-Su Kim for their assistance and advice with the experimental set-up. The authors would also like to thank Iskandar Bahari, Rupesh Kumar and Marco Lucamarini for fruitful discussion on the LM05 protocol and the experiments.

**Appendix A**

It is known that for fully polarized light, one may conveniently uses Jones calculus to represent polarization transformations. In this appendix, we use Jones calculus to describe polarization transformations related to active components mainly Pockels cells used as polarization modulators in the experimental setup starting with source and detector in Bob, followed by the flipper in Alice and finally summarize them in Table 1.

Let us first denote Jones vector for all four linear polarization states from the two basis used in this setup as Vertical, $|V\rangle = \begin{bmatrix} 0 \\ 1 \end{bmatrix}$; Horizontal, $|H\rangle = \begin{bmatrix} 0 \\ 1 \end{bmatrix}$; Diagonal, $|D\rangle = \begin{bmatrix} 1 \\ 1 \end{bmatrix}$ and Anti-diagonal, $|A\rangle = \begin{bmatrix} 1 \\ -1 \end{bmatrix}$. Jones matrix for rotatable half wave plate is given by $\begin{bmatrix} \cos 2\theta & \sin 2\theta \\ \sin 2\theta & -\cos 2\theta \end{bmatrix}$. While Pockels cell not in triggered state is equivalent to identity matrix $I = \begin{bmatrix} 1 & 0 \\ 0 & 1 \end{bmatrix}$, Pockels cell at orientation $\theta$ to optical axis, triggered with half wave voltage is represented as half wave plate with the following Jones matrix $M1 = \begin{bmatrix} \cos 2\theta & \sin 2\theta \\ \sin 2\theta & -\cos 2\theta \end{bmatrix}$. Replacing $\theta = \frac{-\pi}{8}$, in M1 (representing matrix operation of PC1 in Fig 2) becomes $M1 = \begin{bmatrix} 1 & -1 \\ -1 & -1 \end{bmatrix}$. Now, polarization state of photons coming out of the two lasers SRC1 and SRC2 are each represented as $|H\rangle = \begin{bmatrix} 1 \\ 0 \end{bmatrix}$ and $|V\rangle = \begin{bmatrix} 0 \\ 1 \end{bmatrix}$. Solving matrix multiplication against identity matrix I for the case of PC1 not triggered and M1 for the case of triggered will result in all the four linear polarization states as shown in the following equation Eq 11**.

$$I \times |H\rangle = |H\rangle \,;\, M1 \times |H\rangle = |D\rangle \,;\, I \times |V\rangle = |V\rangle \,;\, M1 \times |V\rangle = A \tag{11}$$

The treatment of Jones matrix for the case of the second Pockels cell in Bob (denoted as PC4 in Fig 2) is almost similar to the case of the source except that they are the reverse operation since the function of PC4 is to prepare the same measurement basis as it was prepared at the source. As such, treating M4 (denoted as PC4 in Fig 2) as half wave plate with orientation $\theta = \frac{\pi}{8}$ as it was used in this setup (alternatively one can also use $\theta = \frac{\pi}{8}$ ), results in the following equation Eq 12.

$$I \times |H\rangle = |H\rangle \,;\, M4 \times |D\rangle = |H\rangle \,;\, I \times |V\rangle = |V\rangle \,;\, M4 \times |A\rangle = |V\rangle \tag{12}$$

As previously explained, Alice encodes logical bit to Bob using the flipper denoted as PC2 and PC3 in Fig 2. Note that Alice need to be able to orthogonally rotate any four states from the two basis without priory knowing what the incoming state is and hence the use of two Pockels cells. For this, the first Pockels cell (PC2) is used to orthogonally rotate the rectilinear basis (horizontal and vertical state of polarization) while the second Pockels cell (PC3) is used to orthogonally rotate the diagonal basis. While they are both triggered simultaneously, each operation does not affect the other basis and thus allow a sort of "universal equatorial gate" mentioned in [18]. By setting the orientation of PC2 and PC3 to $\theta = \frac{\pi}{4}$ and $\theta = 0$ respectively, the Jones matrix for PC2 and PC3 in their triggered state are each given by $M2 = \begin{bmatrix} 0 & 1 \\ 1 & 0 \end{bmatrix}$ and $M3 = \begin{bmatrix} 1 & 0 \\ 0 & -1 \end{bmatrix}$. We can further combine the two operations by conducting matrix multiplication for M1 andM2 and obtain $M23 = \begin{bmatrix} 0 & -1 \\ 1 & 0 \end{bmatrix}$. Now, solving matrix multiplication against all four linear polarization states result in the following Eq 13 for flipper not in triggered state (Alice encode bit 0)and Eq 14 for flipper in triggered state (Alice encode bit 1), we obtain the following.

$$I \times |H\rangle = |H\rangle \ ; I \times |A\rangle = |A\rangle \ ; I \times |V\rangle = |V\rangle \ ; I \times |D\rangle = |D\rangle; \qquad (13)$$

$$M23 \times |H\rangle = |V\rangle \ ; M23 \times |A\rangle = |D\rangle \ ; M23 \times |V\rangle = |H\rangle \ ; M23 \times |D\rangle = |A\rangle; \qquad (14)$$

Note that same result can also be obtained if one exchanges the position of PC2 and PC3 as each matrix operation does not interfere polarization state of the other basis. The results of all matrix operations involving all Pockels cells used throughout this experimental setup (PC1, PC2, PC3 and PC4) as described in this section are summarized in table 1. Based on this, the coding of LM05 protocol is embedded in our LabVIEW program running at Bob and Alice. Note that one can also achieve the same objective using different configuration of M1 and M2 in Table 1 combined with the program used to realize the LM05 protocol. To aid in setting up the Pockels cells to achieve the intended polarization transformations, we have extensively made use of our polarimeter from Thorlab.

Table 1 Transformation of polarization states involving Pockels cells used throughout the setup represented using Jones calculus.

| | | Bob | | Alice | | Bob | | |
|---|---|---|---|---|---|---|---|---|
| Quantum states | Laser source | Polarization modulator PC1 | | Flipper PC2 & PC3 | | Polarization modulator PC4 | | Single photon detector |
| | | OFF $\begin{bmatrix}1 & 0\\0 & 1\end{bmatrix}$ | ON $\begin{bmatrix}1 & -1\\-1 & -1\end{bmatrix}$ | OFF $\begin{bmatrix}1 & 0\\0 & 1\end{bmatrix}$ | ON $\begin{bmatrix}0 & -1\\1 & 0\end{bmatrix}$ | OFF $\begin{bmatrix}1 & 0\\0 & 1\end{bmatrix}$ | ON $\begin{bmatrix}1 & 1\\1 & -1\end{bmatrix}$ | |
| Vertical, $\begin{bmatrix}0\\1\end{bmatrix}$ | SRC2 | $\begin{bmatrix}0\\1\end{bmatrix}$ | | $\begin{bmatrix}0\\1\end{bmatrix}$ | | $\begin{bmatrix}0\\1\end{bmatrix}$ | | SPCM2 |
| | | | | | $\begin{bmatrix}1\\0\end{bmatrix}$ | $\begin{bmatrix}1\\0\end{bmatrix}$ | | SPCM1 |
| Diagonal, $\begin{bmatrix}1\\1\end{bmatrix}$ | SRC2 | | $\begin{bmatrix}1\\1\end{bmatrix}$ | $\begin{bmatrix}1\\1\end{bmatrix}$ | | | $\begin{bmatrix}1\\0\end{bmatrix}$ | SPCM1 |
| | | | | | $\begin{bmatrix}1\\-1\end{bmatrix}$ | | $\begin{bmatrix}0\\1\end{bmatrix}$ | SPCM2 |
| Horizontal, $\begin{bmatrix}1\\0\end{bmatrix}$ | SRC1 | $\begin{bmatrix}1\\0\end{bmatrix}$ | | $\begin{bmatrix}1\\0\end{bmatrix}$ | | $\begin{bmatrix}1\\0\end{bmatrix}$ | | SPCM1 |
| | | | | | $\begin{bmatrix}0\\1\end{bmatrix}$ | $\begin{bmatrix}0\\1\end{bmatrix}$ | | SPCM2 |
| Anti-diagonal, $\begin{bmatrix}1\\-1\end{bmatrix}$ | SRC1 | | $\begin{bmatrix}1\\-1\end{bmatrix}$ | $\begin{bmatrix}1\\-1\end{bmatrix}$ | | | $\begin{bmatrix}0\\1\end{bmatrix}$ | SPCM2 |
| | | | | | $\begin{bmatrix}1\\1\end{bmatrix}$ | | $\begin{bmatrix}1\\0\end{bmatrix}$ | SPCM1 |

*SRC1, SRC2, PC1, PC2, PC3, PC4, SPCM1, SPCM2 is from Fig 2.

** Global phase has no influence to polarization state and hence is ignored here.